\documentclass[letterpaper]{article} % DO NOT CHANGE THIS
\usepackage[submission]{aaai25}  % DO NOT CHANGE THIS
\usepackage{times}  % DO NOT CHANGE THIS
\usepackage{helvet}  % DO NOT CHANGE THIS
\usepackage{courier}  % DO NOT CHANGE THIS
\usepackage[hyphens]{url}  % DO NOT CHANGE THIS
\usepackage{graphicx} % DO NOT CHANGE THIS
\def\UrlFont{\rm}  % DO NOT CHANGE THIS
\usepackage{natbib}  % DO NOT CHANGE THIS AND DO NOT ADD ANY OPTIONS TO IT
\usepackage{caption} % DO NOT CHANGE THIS AND DO NOT ADD ANY OPTIONS TO IT
\usepackage{algorithm}
\usepackage{algorithmic}

\usepackage{amsmath}
\usepackage{multirow}
\usepackage{float}
\usepackage{placeins}

\usepackage{xcolor}
\newcommand{\answerYes}[1]{\textcolor{blue}{#1}} 
 
\newcommand{\answerNA}[1]{\textcolor{gray}{#1}}

\usepackage{newfloat}
\usepackage{listings}
\DeclareCaptionStyle{ruled}{labelfont=normalfont,labelsep=colon,strut=off} % DO NOT CHANGE THIS
\floatstyle{ruled}
\newfloat{listing}{tb}{lst}{}
\floatname{listing}{Listing}
\title{Sometimes the Model doth preach: Quantifying Religious Bias in Open LLMs through Demographic Analysis in Asian Nations \\ 
\vspace{0.5em} % Adjust to move closer to the title if necessary
    \small
    {\color{red} \textbf{Warning: This paper contains content that may be potentially offensive or upsetting.}}
}

\author{
    Hari Shankar\textsuperscript{\rm 1}, 
    Vedanta S P\textsuperscript{\rm 2}, 
    Tejas Cavale\textsuperscript{\rm 1}, 
    Ponnurangam Kumaraguru\textsuperscript{\rm 1}, 
    Abhijnan Chakraborthy\textsuperscript{\rm 3}
}
\affiliations{
    \textsuperscript{\rm 1}IIIT Hyderabad\\
    \textsuperscript{\rm 2}IIIT Kottayam\\
    \textsuperscript{\rm 3}IIT Kharagpur
    
}
\begin{document}

\maketitle

\begin{abstract}
% "Novel method that improves"
% Global South - understanding representativeness
Large Language Models (LLMs) are capable of generating opinions and propagating bias unknowingly, originating from unrepresentative and non-diverse data collection. Prior research %works 
has analysed these opinions with respect to the West, %ern, developed populations, 
particularly the United States. However, insights thus produced may not be generalized in %applicability to 
non-Western populations. With the widespread usage of LLM systems by users across several different walks of life, the cultural sensitivity of each generated output is of crucial interest.
Our work proposes a novel method that quantitatively analyzes the opinions generated by LLMs, improving on previous work with regards to extracting the social demographics of the models. Our method measures the distance from an LLM's response to survey respondents, through Hamming Distance, to infer the demographic characteristics reflected in the model's outputs. We evaluate modern, open LLMs such as Llama and Mistral on surveys conducted in various global south countries, with a focus on India and other Asian nations, specifically assessing the model’s performance on surveys related to religious tolerance and identity. Our analysis reveals that most open LLMs match a single homogeneous profile, varying across different countries/territories, which in turn raises questions about the risks of LLMs promoting a hegemonic worldview, and undermining perspectives of different minorities. Our framework may also be useful for future research investigating the complex intersection between training data, model architecture, and the resulting biases reflected in LLM outputs, particularly concerning sensitive topics like religious tolerance and identity.
\end{abstract}

\begin{figure*}[t]
    \centering
    \includegraphics[width=1\linewidth]{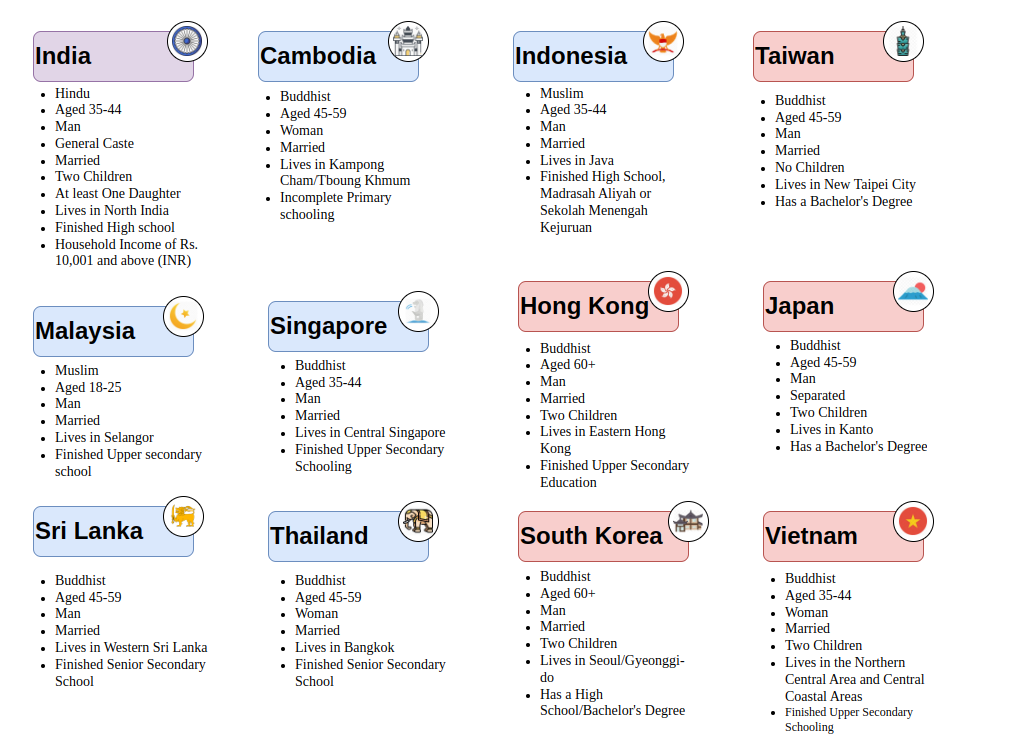}
    \caption{Summarizing Model Profiles. Model Profiles are specific to the surveys used to create them. Colour code for countries: Purple - India. Blue - Countries from Southeast Asia Survey. Red - Countries from East Asia Survey.}
\end{figure*}

\section{Introduction}

%Religion data and trends over the last decade provide several divergent narratives across societies around the world on the evolving importance of religion in everyday life 
Data and trends over the past decade reveal contrasting narratives about the evolving role of religion in societies worldwide~\cite{storm2024religion}.
%Developed nations, such as the United States, Australia, New Zealand and various countries of Western Europe suggest a decline in its importance 
In the United States, Australia, New Zealand, and various Western European countries, there has been a notable decline in the perceived importance of religion~\cite{pew2022keyFindings}. For instance, the proportion of self-reported Christians in the United States dropped from $78\%$ to $63\%$ within the last decade%, with a reduction of self-reported Christians from 78\% to 63\% over the last decade 
~\cite{pew2021USAdults}. In contrast, data from %countries in the Global South, including India and Indonesia, 
South Asia suggest a stronger alignment between religion and politics. In these regions, voters often favor leaders who share their religious beliefs and advocate for the interests of their religious communities, underscoring the enduring influence of religion in shaping societal and political preferences~\cite{pew2024leaderbeliefs}. 
Similarly, surveys conducted across Europe highlight a divide in social attitudes. Respondents in Central and Eastern Europe were found to be less accepting of individuals from Jewish or Muslim backgrounds as part of their family compared to those in Western Europe, signaling regional differences in religious tolerance~\cite{pew2018EastWestEurope}.

%Meanwhile, surveys held in several countries across Europe found that most respondents belonging to Central and Eastern Europe were less likely to accept individuals of Jewish or Muslim backgrounds as a part of their family, as compared to their counterparts in Western Europe \cite{pew2018EastWestEurope}. 
%In addition to this, polls held in countries such as India, Indonesia and other nations generally considered a part of a Global South suggest that most voters in these countries prefer leaders that have the same religious beliefs as their own, and stand up for the people of their religion \cite{pew2024leaderbeliefs}. 

The rise of Large Language Model (LLM) powered AI chatbots like ChatGPT and Claude\footnote{\url{https://chatgpt.com}; \url{https://claude.ai}} has significantly transformed the way people engage with the Internet and seek information, including topics related to religion. Traditionally, individuals relied on search engines and resultant websites to explore religious teachings and discussions. Now, many turn to conversational AI for direct answers about faith, practices, and even moral dilemma~\cite{ai_faith_ethics}. This shift reflects the broader trends in religious attitudes discussed earlier, as societies worldwide navigate differing perspectives on religion's role in everyday life. In regions where religion is deeply intertwined with identity and governance, an LLM's ability to provide nuanced and culturally sensitive responses may hold particular significance. On the other hand, in secularizing societies, an LLM could become a tool to foster interfaith understanding or addressing questions from an increasingly diverse population~\cite{venkit2024generative}.

%The rise of AI chatbots such as ChatGPT\footnote{}, Claude has radically changed how people interact with the Internet and seek information online, as traditionally people used search engines to find religious teachings and discussions, many now turn to conversational AI for direct answers about faith and practices.
%However, this shift raises important questions. How well can AI systems understand and convey nuanced religious concepts? How unbiased are they regarding the sentiments of individuals or do they have an opinion of their own? \cite{venkit2024generative}

However, this evolution raises critical questions: {\it How effectively can AI systems understand and convey the complexities of religious concepts}, which are often deeply personal and context-specific? Are these systems capable of remaining unbiased, or {\it do they inadvertently reflect societal or cultural biases embedded in their training data}? In this paper, we analyze how various LLMs respond to questions pertaining to religious identity. We seek to determine whether any model responds in a manner that does not reflect the religious sensibilities of the population. We consider Open LLMs such as Llama \cite{touvron2023llama} and Mistral \cite{jiang2023mistral}, as opposed to other closed LLMs such as ChatGPT or Claude. An Open LLM is a language model with publicly accessible architecture, training code, and often pre-trained weights, enabling anyone to use, modify, and share it. This openness is valuable as it allows developers to customize or enhance the model more easily compared to closed LLMs.

Our work aims to profile different LLMs by different social demographics. Our analysis is the first step in understanding the model’s biases on the topic, which may then be used to create safety guard-rails to mitigate the same. We propose a novel method to determine the demographics reflected by the model that recognizes the interconnected nature of different demographic variables such as Age and Education Level, etc. We release the codebase for our experiments and results publicly on GitHub.\footnote{\url{https://github.com/HariShankar08/LLMOpinions}} In addition to this, we also examine the effectiveness of instructing LLMs within prompts to respond as though they belonged to any particular group. Our results indicate no significant variation in the model’s profile by virtue of this method.

% Risks of Steering is a To-Do in Disucssion.

\section{Background and Related Work}

\subsection{Religion and the Internet}

Usage of the Internet as an avenue of proliferating religion has been examined in the past. Works conducted in predominantly Muslim societies such as Indonesia suggest that individuals belonging to younger generations are increasingly turning to social media to learn about religion \cite{thoriquttyas2021religious}. The COVID-19 pandemic also brought about changes to the ways people approached worship; for example, on the 27th of March, 2020, Pope Francis gave a special \textit{Urbi et Orbi} blessing in an empty St. Peter's Square, accessible to the public only through television or Internet streaming \cite{watkins2020pope}. Applications such as Zoom, Microsoft Teams and Google Meet have also been used by different religious institutions as a platform to provide their services to the public. Additionally, various software systems such as ParishSoft\footnote{https://www.parishsoft.com/religious-education/}, have been developed over the years to aid in religious teaching \cite{hamdelsayed2016islamic}, which have correspondingly evolved with the advent of conversational AI, and LLMs. More recent studies have examined the usage of LLMs such as ChatGPT in aiding in biblical studies \cite{chrostowski2024chatgpt}. While such systems may stand the risk of providing erroneous information, we observe that on average, LLMs provide satisfactory responses to the user's queries. 

\subsection{Human-LLM interaction}

The term LLM refers to Large Language Models, which are AI systems trained on large amounts of textual data, generally sourced from the Internet and other publicly available resources. Most common LLMs build off the generative pre-trained transformer (GPT) architecture \cite{radford2018improving}, and are capable of generating coherent, natural language text. Several significant works in the past have been conducted to evaluate the performance of these models on tasks such as language understanding \cite{hendryckstest2021}. Our work instead focuses on situations where LLMs may be used to aid the human learning and decision-making processes. ChatGPT has been documented to aid users in several different aspects of day-to-day life, such as mental health \cite{stade2024large}, and relationship advice \cite{Hou_Leach_Huang_2024}. Our work seeks to examine LLMs under a similar lens, by evaluating how these systems may be used to aid a user in exploring their religious identity.    

\subsection{Profiling LLMs}

Our study takes inspiration from previous work by Santurkar et al. (2023) which profiled LLMs on the basis of various social demographics such as political ideology, household income, and education level \cite{santurkar2023whose}. The profiling was done by prompting different models to respond to public survey questionnaires, following which the model's response was compared to the actual survey respondents belonging to the different demographic groups. The study reported that LLMs such as GPT 3.5 (text-ada and text-davinci models) responded to the selected survey questionnaires similar to respondents who were liberal, highly educated and belonging to higher income groups in the United States, whereas other models (ada and davinci) responded similar to moderate-to-conservative individuals belonging to lower income strata. Our work profiles modern open LLMs based on surveys held in India and various countries of Eastern and South-Eastern Asia, where societies are noted to emphasize the importance of religion, collectivism and communal ties as opposed to the more individualistic nature of western nations \cite{Maguire2017}.

\subsection{The Global South}

The Global South is a group of countries characterized by shared experiences of colonialism and economic development status, not strictly defined by geography. The group is also frequently associated with the G77 group of nations, including nations such as India, Indonesia and Vietnam. Historically, many of the nations belonging to the Global South have been marginalized in global politics, however, they have experienced a significant shift in wealth and political visibility in recent decades. The World Bank has acknowledged a `shift in wealth' from the North Atlantic to the Asia-Pacific region, challenging conventional notions of economic power distribution \cite{worldbank2024eap}.

\section{Dataset}

\begin{figure*}[!t]
    \centering
    \includegraphics[width=1\linewidth]{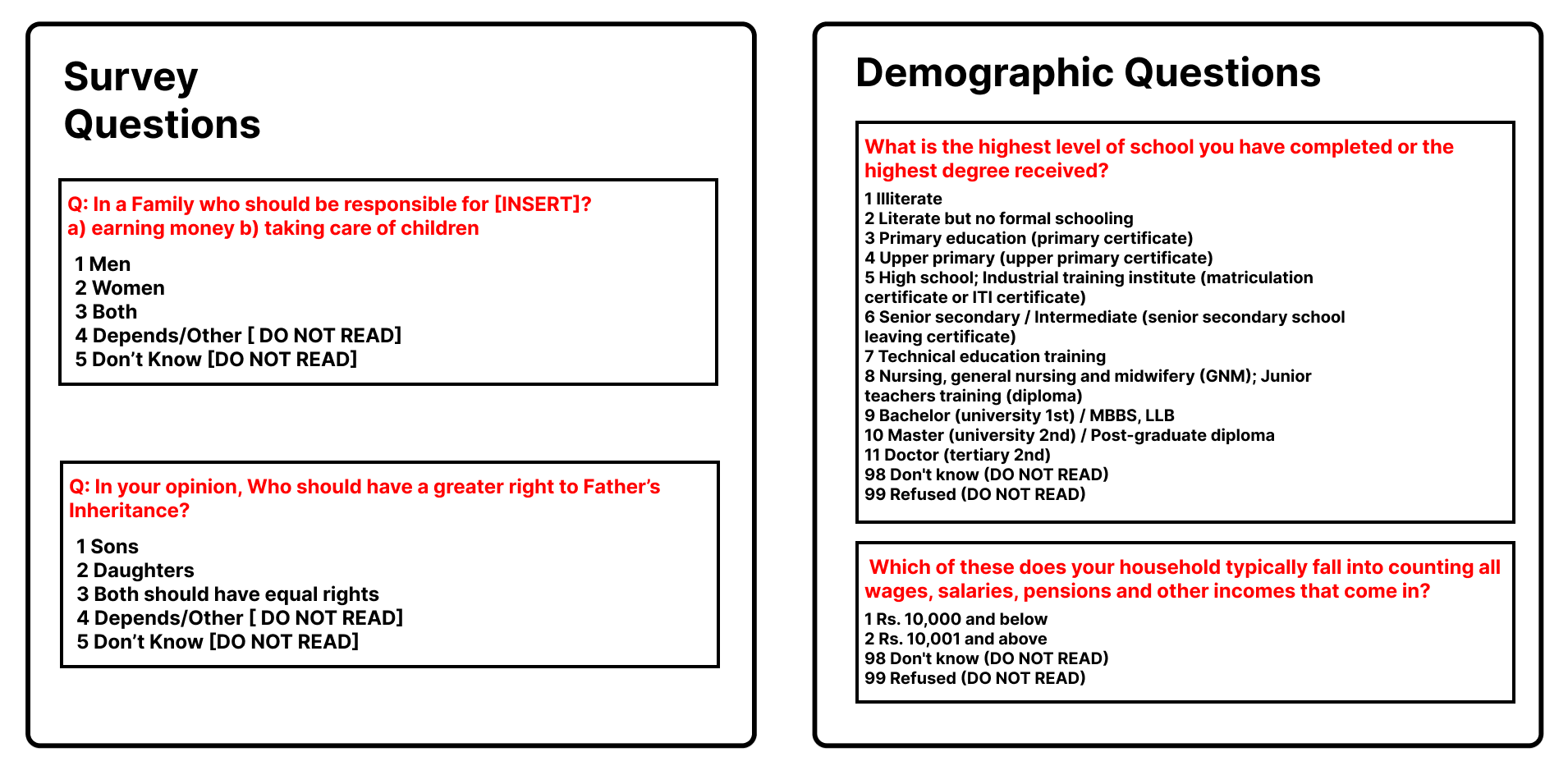}
    \caption{Sample Survey Questions from the Indian Survey Dataset. The left table presents opinion-based questions focused on family responsibilities and inheritance rights. The right table includes few demographic questions, such as education level and household income.}
    \label{fig:questions}
\end{figure*}

For our study, we used three publicly available survey data from the Pew Research Center \cite{indiaSurvey, EASurvey,SEASurvey}, each of which offers an extensive exploration of religious beliefs and practices across different Asian societies. The datasets consist of several questionnaire and the responses %collected from the survey, as well as the  used by 
provided by the survey respondents. 

The India Survey Dataset surveys the Indian public about religious beliefs and practices, religious identity, nationalism, and tolerance in Indian society. In addition to the India Survey, we also use two other datasets which focus on various aspects of the survey public, including demographics, socio-economic factors, and public opinion. The East Asian Societies Dataset, which we henceforth refer to as the EA dataset, consists of survey responses from five different territories: Japan (JPN), Hong Kong (HKG), South Korea (KOR), Taiwan (TWN) and Vietnam (VNM). We also use survey responses from South and South-East Asia Datasets, henceforth referred to as the SEA dataset, from the following nations: Indonesia (IDN), Malaysia (MYS), Singapore (SGP), Sri Lanka (LKA) and Thailand (THA).

Each dataset provides a CSV file where the rows represent individual survey responses, and the columns correspond to the survey questions and variables used to characterize each respondent. In addition to this, an XML metadata file is also provided which maps the options of each survey question to the value used to represent the same option in the responses file.

\begin{table}[t]
\centering
\begin{tabular}{|l|l|}
\hline
\textbf{Dataset} & \textbf{N. Respondents} \\ \hline
India            & 29,999                   \\ \hline
EA               & 10,390                   \\
(of which) JPN   & 1,742                    \\
(of which) HKG   & 2,012                    \\
(of which) KOR   & 2,104                    \\
(of which) TWN   & 2,277                    \\
(of which) VNM   & 2,255                    \\ \hline
SEA              & 13,122                   \\
(of which) IDN   & 2,571                    \\
(of which) MYS   & 1,999                    \\
(of which) SGP   & 2,036                    \\
(of which) KHM   & 1,502                    \\
(of which) LKA   & 2,510                    \\
(of which) THA   & 2,504                    \\ \hline
\end{tabular}
\caption{Breakdown of respondents by region: East Asia (EA), including Japan (JPN), Hong Kong (HKG), South Korea (KOR), Taiwan (TWN), and Vietnam (VNM); Southeast Asia (SEA), including Indonesia (IDN), Malaysia (MYS), Singapore (SGP), Cambodia (KHM), Sri Lanka (LKA), and Thailand (THA).}
\label{tbl:respondents}
\end{table}

Each respondent is identified by a unique numeric value, represented in the responses files as the ``QRID" of the respondent. Each survey has been designed so that each respondent also provides demographic information such as age, gender, etc.

The Indian Survey dataset has a total of 308 columns, representing survey questions as well as additional variables such as the region and ISCED-97 \cite{unesco2003international}. These additional variables are directly recorded by interviewers. Similarly, the surveys held in the East, South and South-East Asian nations each contain a total of 174 columns.

Each survey consists of different demographic variables which we use to profile each LLM, of these, we note the following five common variables: Religion, Gender, Age, Marital Status, Region and Education Level. The EA Survey Dataset additionally provides the Number of Children of each respondent. In addition to the five variables previously listed, we use the Urbanicity of the region to profile models in the SEA Survey countries. For the India Survey, we are provided with the following additional variables: Caste, Number of Daughters, and Monthly Household Income.  

\subsubsection{Language of the Surveys} 

In the study carried out by the members of Pew Research Center, each survey questionnaire was developed in English, following which the questionnaires were translated and verified by professional linguists having a native proficiency in the languages. For our study, we directly use the questionnaire in English, without any translation steps. Translation has been omitted since we do not have the means to reliably verify the correctness of machine-translated prompts in each of the target languages. Additionally, as LLMs are mostly trained on English textual data, we may expect to find more biases associated with English-speaking regions of the world, as compared to other less represented regions.

\section{Methodology}

\subsection{The Model Response}

We use the term Model Response to refer to the collection of all answers provided by the model for each question of the corresponding survey questionnaire. For each survey, we first determine the questions that directly probe into the demographic of the respondent. This step is manual, through which we compile two sets of questions, the Survey Questions and the Demographic Questions. We provide some examples of these questions in Figure \ref{fig:questions}.

Although the survey questionnaire instructed not to read options for refusal or when the respondent did not know the answer, we omit such instructions when constructing the prompt for the model. This was achieved using regular expressions (regex). This step ensures that all options are explicitly provided to the LLM.

We prompt each model on the selected Survey Questions, and select the option having the highest token probability for our answer. However, to account for Prompt Sensitivity potentially leading to different answers for the same question \cite{anagnostidis2024susceptible}, we additionally opt to paraphrase the survey questions using Parrot \cite{prithivida2021parrot}. Each prompt is run on five random seeds for ablation purposes. Our system then
selects the most frequent answer across all candidate answers, and reports this as the model's answer to a particular question.

\subsection{Profiling the Model}

\begin{figure}[t]
    \centering
    \includegraphics[width=1\linewidth]{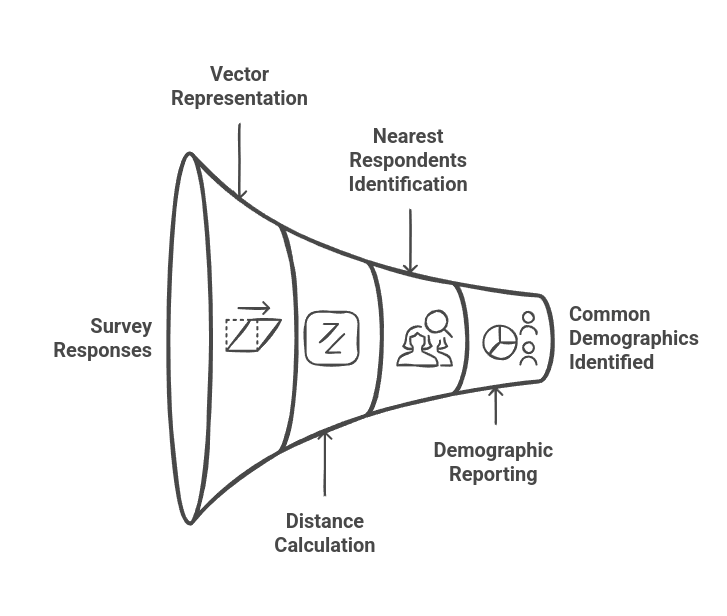}
    \caption{System Overview for Model Profiling. Distance is computed by representing both Model and Survey Response as a Vector, following which the Hamming Distance metric is calculated between the two vectors.}
    \label{fig:model-profile}
\end{figure}

The Hamming Distance is a metric originally defined in Information Theory, measuring the number of positions at which corresponding members of equal dimension vectors are different \cite{hamming1950error}. The metric is also used in the domain of Machine Learning, to compare vectors containing categorical variables. Our approach is largely inspired of the K-Nearest algorithm. 

\begin{equation}
\label{eqn:hamming}
d_H(\mathbf{A}, \mathbf{B}) = \frac{1}{n} \sum_{i=1}^n \delta(A_i, B_i), \quad \text{where } n = |\mathbf{A}| = |\mathbf{B}|
\end{equation}

\begin{equation}
\label{eqn:indicator}
\delta(A_i, B_i) = 
\begin{cases} 
1, & \text{if } A_i \neq B_i \\
0, & \text{if } A_i = B_i
\end{cases}
\end{equation}

The surveys also contain questions that are not applicable to all groups within the survey public. In the case of such questions, where a respondent has not been asked a particular question, a blank value is provided in the corresponding column in the dataset files. Before computing Hamming Distance, we drop any column which contains blank values.  

Each response vector, which now strictly consists of numerical values that represent the option selected, can be considered as a vector of categorical variables. This is applicable as the options here do not possess any intrinsic quality with respect to ordering. We encode all vectors using the One-Hot Encoding algorithm, and compute the Hamming Distance as defined in the equations \ref{eqn:hamming} and \ref{eqn:indicator}. As lower Hamming Distance suggests more similar vectors, we in turn sort the survey respondents by increasing values of the metric to get the closest respondents to the Model Response. 

Our method enables us to extract all demographic variables for the model using a single distance value. This is in contrast to previous methods \cite{santurkar2023whose, durmus2023towards} which involved the computation of a closeness metric for each individual demographic variable. However, such an approach risks over-representation, and does not consider the interconnected nature of the demographic variables. The proposed metric allows the model's response to be appropriately placed within clusters defined by the different demographic variables, which in turn constitutes the model's profile. We summarize our method in Figure \ref{fig:model-profile}.

We note that our approach would result in the model profiles reflecting sampling biases present among the survey respondents. As such, the model profiles may not be extended beyond the scope of the surveys used to create them. 

\subsection{Steering}

\begin{figure}[t!]
    \centering
    \includegraphics[width=1\linewidth]{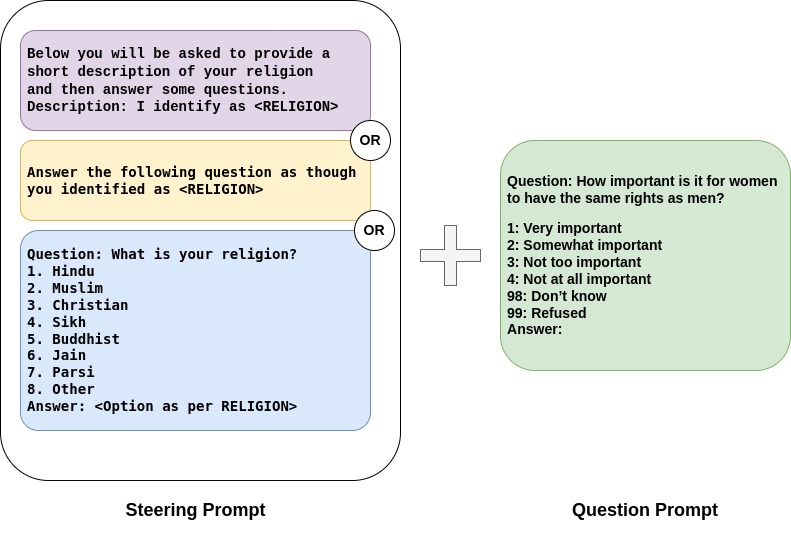}
    \caption{Detailing the different steering prompts used in our experiments.}
    \label{fig:steering-explainer}
\end{figure}

Steering refers to the various methods used to make LLMs respond more similarly to any specific group, which we shall henceforth refer to as the steering group. Some methods documented in previous works include fine-tuning \cite{jiang2022communitylmprobingpartisanworldviews} and more generic data-driven LLM augmentation \cite{li2024steerabilitylargelanguagemodels} which use data from a cohort of individuals manifesting similar views across specific inquiries, to steer towards the steering group.

We limit our discussion to the effectiveness of zero-shot steering capabilities of LLMs, in the absence of any additional training or augmenting methods. In such a setting, instructions to steer the model towards a particular steering group are provided within the prompt itself. We opt for the following methods to steer the model:

\begin{enumerate}
    \item Through a multiple-choice question, whose answer is also provided in the prompt.
    \item Through a description of the steering group to portray.
    \item Through an explicit instruction to answer the question as though it belonged to a particular demographic group. 
\end{enumerate}

The exact prompts used in our study have been provided in Figure \ref{fig:steering-explainer}. Steering prompts are prepended to each question prompt before running inferences on the model. We examine the effectiveness of steering a LLM towards different religions, using responses from the Indian Survey dataset. We examine the average Hamming Distance between the model response without any steering, and respondent belonging to each religion. We then compute the same values after steering the model. 

\subsection{Models Used in the Study}

% Please add the following required packages to your document preamble:
% \usepackage{multirow}
\begin{table}[t]
\centering
\begin{tabular}{|l|c|c|}
\hline
\textbf{Model}             & \textbf{Configuration} & \textbf{Short-hand} \\ \hline
\multirow{2}{*}{Llama 3.2} & 3B                     & L3B                 \\ \cline{2-3} 
                           & 3B Instruct            & L3BIt               \\ \hline
\multirow{2}{*}{Llama 3.1} & 8B                     & L8B                 \\ \cline{2-3} 
                           & 8B Instruct            & L8BIt               \\ \hline
\multirow{4}{*}{Mistral}   & 7B v0.1                & M7Bv1               \\ \cline{2-3} 
                           & 7B v0.1 Instruct       & M7Bv1It             \\ \cline{2-3} 
                           & 7B v0.3                & M7Bv3               \\ \cline{2-3} 
                           & 7B v0.3 Instruct       & M7Bv3It             \\ \hline
\multirow{2}{*}{Gemma 2}   & 9B                     & G9B                 \\ \cline{2-3} 
                           & 9B Instruct            & G9BIt               \\ \hline
\end{tabular}
\caption{Different Open LLMs used in our study, the configuration used per model (Number of Parameters), and short-hand used to refer to the model in the model profile chart (Figure \ref{fig:ind-profiles}).}
\label{tbl:models}
\end{table}

We list the models evaluated in Table \ref{tbl:models}. Our study examines the profiles of major modern open LLMs. All models are run on an Apple Macbook Pro, using the M4 chip and the MPS backend provided by the HuggingFace library\footnote{https://huggingface.co/docs/diffusers/en/optimization/mps}. As our approach only accesses the logits of the model to extract the most probable token, we are not required to modify any hyperparameters generally associated with the text generation process. However, to ensure reproducibility of results, we initialize our model runs by setting all random seeds before any inferences.

\section{Results}

We examine the closest respondents to the Model Response to extract a profile for the model. For our study we define the model's profile as a collection of the modal demographic attributes among these respondents. In the case of the India survey, this is done by considering the top 1000 closest respondents, due to the significantly larger number of survey respondents. For each country in the EA and SEA survey, we instead consider the top 100 closest respondents.  

\subsection{India}

\begin{figure*}[h]
    \centering
    \includegraphics[width=0.9\linewidth]{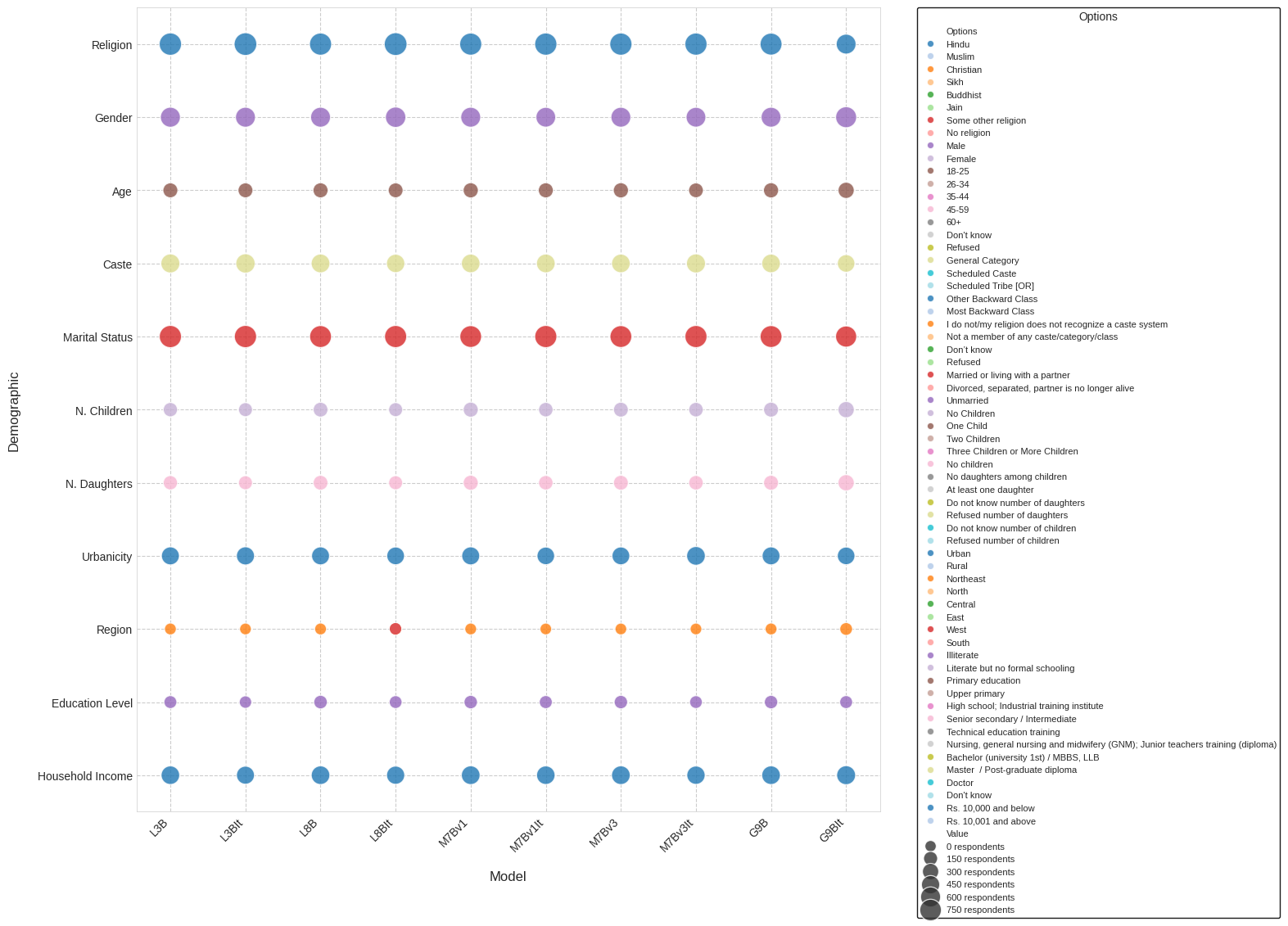}
    \caption{Model Profiles - India. Comparing Model Responses with top 1,000 closest Survey Respondents. LLMs primarily converge on a single demographic profile (e.g., married Hindu males, 35-44, rural Northern India, low income, high school), with minor regional or educational variations in some models.}
    \label{fig:ind-profiles}
\end{figure*}

We find that most selected LLMs result in a single, homogenized profile, with exceptions to this statement usually only differing by at most two different demographic variables. Figure \ref{fig:ind-profiles} further indicates the uniformity of responses across different models. We continue to observe this trend even in the EA and SEA surveys.

With regards to the Indian survey dataset, we report the following profile: models indicate strong correspondence with respondents aged 35-44 years who are either married or living with a partner. We note significant alignment with Hindu male respondents from the General caste. Among the top 1,000 respondents, the majority report monthly household incomes above Rs. 10,000 (INR), with predominantly a high-school level of education.

Geographically, the models align with respondents from rural settlements in northern India. However, exceptions include Llama 3.1 Instruct and Gemma 2 Instruct, which show stronger alignment to respondents from western India. Additionally, Llama 3.1 Instruct exhibits a distinctive pattern in educational alignment, corresponding more closely with respondents holding primary education certificates, contrasting other models in the analysis. We provide the model profiles in detail in Figure \ref{fig:ind-profiles}.

\subsection{East Asia}

For the subsequent surveys, we extract the model's profile from the top 100 closest respondents to the Model Response. This is carried out for each territory to provide individual profiles of the model. We observe that across most territories, with the exception of Vietnam, models are closely aligned with Married, Male individuals having 2 children. In the case of Vietnam, however, we only observe a variation in the Gender demographic, with closer alignment to Women respondents. Detailed profiles for each country have been provided in the GitHub repository.\footnote{\url{https://github.com/HariShankar08/LLMOpinions}}

\subsubsection{Japan}

The model profiles are closely aligned with Married, Male respondents, aged 45-59 years old. We observe more closer alignment to more educated individuals, here most models reflect the opinions of University-educated respondents. Models are aligned to be of predominantly Buddhist disposition.

%\begin{figure*}[h]
%    \centering
%    \includegraphics[width=1\linewidth]{EA-modelProfiles/JPN.jpg}
%    \caption{Model Profiles - Japan. Comparing Model Responses with Top 100 closest Survey Respondents.}
%    \label{fig:jpn-profiles}
% \end{figure*}

\subsubsection{Hong Kong}

Model Profiles exhibit a notable alignment with individuals following Buddhism, predominantly from the Eastern Region of Hong Kong. The profiles also clearly tend towards individuals who have achieved at least upper secondary education with exceptions of Llama 3.2 3B Instruct and Llama 3.1 8B which reflect an undergraduate education level.

\subsubsection{South Korea}

Model Profiles reveal distinct characteristics when compared to the general survey public. A notable alignment is observed between the models and individuals who identify as Buddhist, contrasting sharply with the majority of the public, who reported having no religious affiliation. Moreover, The models predominantly mimic individuals older than 60 years of age, who are residing in Seoul or Gyeonggi-do, regions that represent significant urban and cultural hubs in South Korea.

\subsubsection{Taiwan}

Models are largely aligned with individuals of Buddhist faith, aged between 45 and 59 years, living in New Taipei City. This geographic concentration suggests that the model captures patterns characteristic of urban regions.
The educational background shows a strong alignment with Bachelor's degree holders, while gender distribution remains balanced between males and females. These individuals are typically married with no children.

\subsubsection{Vietnam}

Models are largely aligned with individuals of Buddhist faith, aged between 35 and 44 years, living in the Northern Central Area or Central Coastal Areas, having a Bachelors degree. Profiles observed in Vietnam also show a higher alignment to women, in contrast to our findings in other territories within the EA survey, however, this is due to the larger number of women respondents as compared to men in the survey public.  

\subsection{South East Asia}

Similar to the EA Survey, we refer to the top 100 respondents to extract demographics of the model. We observe that across most territories, models are closely aligned with Married, Male individuals.

\subsubsection{Indonesia}

Models are closely aligned with respondents of an age group of 35-44 years old. We observe maximum alignment to individuals of a high-school level of education. Models are closely aligned to Islamic practices. There is no significant variation across different model families. We observe a more equitable distribution, as compared to findings from the Indian survey, with respect to Urbanicity.

Most of the closest respondents, across all models, belong to the Java region of Indonesia. The concentration of respondents from this region suggests the predominance of publicly available online data originating from this region, potentially leading to regional bias in the dataset. 

% \begin{figure*}
%    \centering
%    \includegraphics[width=1\linewidth] {SEA_modelProfiles/IDN.jpg}
%    \caption{Model Profiles - Indonesia}
%    \label{fig:idn-profiles}
% \end{figure*}

\subsubsection{Malaysia}

Most models exhibit strong alignment with individuals from Malaysia who are male, married, and have an educational qualification of Upper Secondary School or higher. This trend suggests that the models are tuned to reflect the perspectives of individuals with a moderate to high level of formal education. The models demonstrate a notable affinity toward individuals practicing Islam. This religious alignment suggests that the data used for model training likely emphasizes cultural or societal norms prevalent within Islamic communities.

\subsubsection{Cambodia}

The Model closely aligns with individuals practicing Buddhism primarily from Rural regions. Interestingly, it reflects a demographic with incomplete Primary Schooling, finished up to Grade 5, indicating a focus on individuals with lower levels of formal education.
Unlike other profiles, this model does not exhibit consistent patterns in terms of age or specific regions within the country. 

\subsubsection{Singapore}

The models provide a response closest to individuals having a Bachelor's Degree, closely aligned to the Buddhist faith, then followed by Hinduism. We note that Singapore, being a more developed nation than its counterparts in the SEA survey, does not consider the urbanicity variable, with all respondents live in Urban regions.  

\subsubsection{Sri Lanka}
The model exhibits a strong alignment with married individuals who follow Buddhism, reside in rural regions, and have completed their education up to the senior secondary level. This alignment underscores the model’s inclination toward a specific demographic that combines religious, marital, and educational attributes, with a pronounced focus on rural populations.
The model does not show a distinct preference or uniformity concerning gender or specific regions within the country.

\subsubsection{Thailand}

We observe an equitable distribution of Urbanicity across models. The models align closely with individuals aged 45-59 years, with an education level of upper secondary schooling. Some outliers include L8BIt and G9BIt which are more closely aligned towards individuals of a primary schooling level.

\subsection{Examining Steering}

\begin{figure}[t]
    \centering
    \includegraphics[width=1\linewidth]{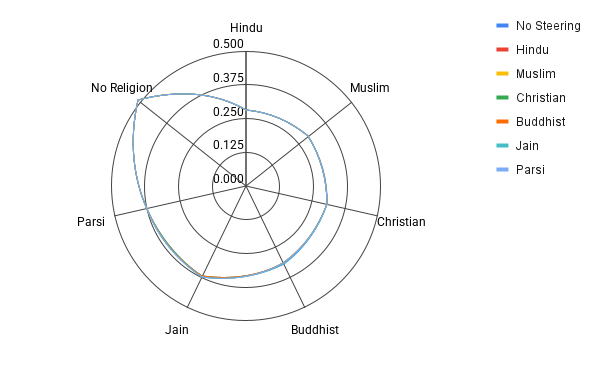}
    \caption{Radar Plot for steering Llama 3.2 3B (L3B): Radius represents average Hamming Distance to different religions. We note no significant variation between the plot without steering, and plots with steering towards religions.}
    \label{fig:rad-L3B}
\end{figure}

\begin{figure}[t]
    \centering
    \includegraphics[width=1\linewidth]{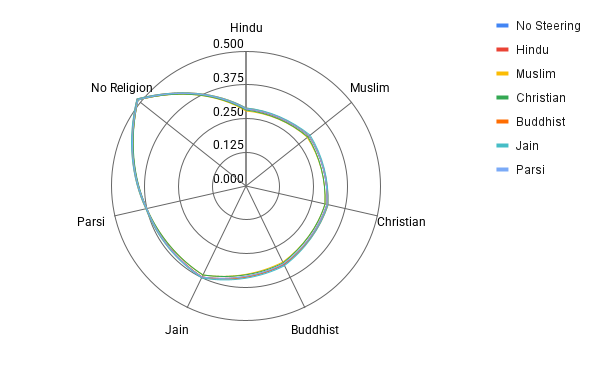}
    \caption{Radar Plot for steering Llama 3.2 3B Instruct (L3BIt): Radius represents average Hamming Distance to different religions. We note the same pattern as in steering L3B, indicating no additional benefit in selecting the instruction tuned variant for steering.}
    \label{fig:rad-L3BIt}
\end{figure}

We use the Llama 3.2 models in our subsequent experiments. Steering is done on the basis of religion. We use the Indian survey, and first generate the model response from the survey questions. As discussed previously, we use the Parrot paraphrasing model to generate multiple versions of the same question, as a precaution for prompt sensitivity. Similarly, in our steering experiments, we use three copies of each question prompt, including paraphrased versions, each of which uses a different steering prompt. The final answer of each question is obtained through the model answer from all candidate answers from each prompt used.

We then calculate the Hamming distance between the Model Response and each survey respondent. Aggregating the survey respondents by the different religions, we examine the average Hamming distance between the Response and each religion. This experimental setup is first carried out in the absence of steering, and then repeated with steering towards each religion. We plot our findings for L3B and L3BIt in Figures \ref{fig:rad-L3B} and \ref{fig:rad-L3BIt} respectively.

We note that on average, the model does not change significantly with respect to the steering religions, even after providing additional context. On examining generated texts from these prompts, which detailed the model's reasoning for picking any particular option, we note that the model emulates the steering group through language usage; utilizing Arabic salutations and honorific terms while portraying a Muslim respondent, etc. However, we find that the opinions, by and large, have not changed significantly. We observe this both in the case of L3B and L3BIt, indicating that additional Instruction tuning does not aid in making the LLM more steerable.

\section{Discussion}

We note that the LLMs provide opinions on religious identity that are largely appropriate in the different nations and territories used in the study. The opinions reflect respondents corresponding to the different demographics as determined in the model profile. We observe minimal variation from this profile across the different families of LLMs used as a part of our study.

We have, in this work, presented a novel approach to determine the social alignment of LLMs and applied the same to public surveys in India and other non-Western countries. Our work also examines the effectiveness of steering by the means of additional context in a prompt, and finds no significant improvement with respect to reflecting opinions of actual members of the particular steering group. The organizing agency, Pew Research Center, bears no responsibility in the analyses or interpretations of the data presented in this work.

\subsection{Data: An Echo Chamber}

Data is the food for models which drive decision making in diverse domains. The quality of an LLM directly depends on the data used to train it. The most common ways to currently collect data include surveys and online data due to their scalability and efficiency but these approaches are not perfect and are plagued by significant limitations, such as under-representation of certain groups and disloyal responses. This leads to models which do not represent a diverse set of societies rather a privileged part of society who have the access \cite{pew2015}. 

\subsection{Social Desirability Bias}
Survey questions asking about taboo topics such as sexual activities, illegal
behaviour such as social fraud, or unsocial attitudes such as racism, often generate inaccurate survey estimates which are distorted by social desirability bias. Due to self-presentation concerns, survey respondents under-report socially undesirable activities and over-report
socially desirable ones \cite{krumpal2013}. In India, for example, respondents might deny a preference for male children, even though cultural practices indicate otherwise. This discrepancy distorts the training data for LLMs, making them less representative of societal truths.

\subsection{Variation through Instruction Tuning}

We note that Instruction tuning does not significantly change the profile of the model itself. This contrasts to what had been observed in previous works \cite{santurkar2023whose} - where additional tuning had resulted in the shift from individuals who were moderate/conservative with low income to those who were liberal and high income. In the cited work, this had been done through RLHF methods \cite{stiennon2020learning}, which had resulted in the model being more susceptible of the human opinions involved in the training. Instruction tuning \cite{wei2021finetuned} in contrast, does not involve any human feedback, and optimizes the imitation of a definite, usually human-written response, which does not result in such a divergence.

\subsection{Amplifying Bias}

Our findings indicate that most LLMs tend to represent specific demographics, a consequence of training on datasets that are often derived from one another. This phenomenon suggests that raw, real-world data is inherently incomplete and lacks sufficient diversity \cite{rodriguez2019defense}. The combination of the models' homogeneous profiles, as well as the ineffectiveness of zero-shot steering amplifies the risk of bias towards these demographics, as AI generated content continues to enter spaces of the Internet. To address this issue, potential solutions include augmenting datasets with synthetic data to introduce greater diversity or employing machine unlearning techniques to mitigate biases inherited from training data. These approaches could help reduce demographic skew and improve the fairness and accuracy of AI-generated survey responses.

\subsection{Shortcomings of Model Profiles}

The model profiles provided must be interpreted with several important caveats: (1) We only use the profiles to indicate the similarities between different LLMs. (2) The profiles inherently reflect sampling biases among survey respondents. Profiles cannot be generalized beyond the respective surveys that were used to create them.  The model profiles must not be misused for stereotyping or discrimination against any particular group of society.

In addition to this, we also face the issue of Alignment Faking; like humans who fake their opinions when put through surveys due to social bias and stigma, LLMs can also fake their alignments \cite{hubinger2024alignment}. This is done by selectively following training objectives and guidelines during supervised sessions, while reverting to undesirable and biased behaviour in cases when the model knows it is not being monitored. This phenomenon highlights the true challenge of alignment, which is the genuine adoption of diversity of opinions rather than just optimizing for surface-level compliance.

\subsection{Limitations and Future Work}

The method described in the paper uses all questions provided in the survey, with minimal manual inputs, in determining the Demographic Questions. However, such an approach underscores an assumption that all questions in the survey have equal relevance in determining the model's profile. A more thorough analysis with respect to the correlations between different questions may help in more effectively determine a model's profile; however, this approach may not be as easily generalizable for surveys across different domains.

We may observe different model profiles when the model is prompted in the different languages used while conducting the survey. This is due to the fact that each corpus used to train LLMs in the target languages may propagate its own biases which are distinct from corpora in other languages such as English.

We may also examine the correlation between generated opinions and content on the Internet from different regions, countries and territories, which constitutes the training data of LLMs. For example, we may examine content on Wikipedia or Reddit to determine correlations between the text and opinions portrayed by the LLM.  

Future works may also choose to examine the effects of features such as continuous access to Internet data beyond the model's training cutoff date, which are typical of more recent proprietary AI systems. We may also use our methodology to evaluate the effectiveness of other steering methods, such as fine-tuning and data augmentation.

\bibliography{citations}

\end{document}